\documentclass[10pt,amsmath,amssymb,nofootinbib,twoside,twocolumn,superscriptaddress,floats,floatfix,aps,prd,preprintnumbers]{revtex4-2}
\usepackage[british]{babel}
\usepackage{xcolor}
\usepackage{booktabs}
\usepackage{latexsym}
\usepackage{amssymb}
\usepackage{amsmath}
\usepackage{graphicx}
\usepackage{tabularx}
\usepackage{cancel}
\usepackage{hyperref}
\hypersetup{
    colorlinks=true,
    linkcolor=black,
    citecolor=blue,
    filecolor=black,
    urlcolor=blue
}
\usepackage[capitalise]{cleveref}
\usepackage[utf8]{inputenc}
\usepackage[normalem]{ulem}
\def\beq#1\eeq{\begin{align}#1\end{align}}

\newcommand{\mms}{M_{\mbox{\tiny MS}}} 

\newcommand{\dd}{\text{d}}

\newcommand{\K}{\mathcal{K}}
\newcommand{\lb}{\label}

\begin{document}

\newcommand{\UniCa}{\affiliation{Dipartimento di Fisica, Universit\`a di Cagliari, Cittadella Universitaria, 09042 Monserrato, Italy}}
\newcommand{\INFNCa}{\affiliation{INFN, Sezione di Cagliari, Cittadella Universitaria, 09042 Monserrato, Italy}}
\newcommand{\CosmoUfes}{\affiliation{Departamento de Física \& Núcleo de Astrofísica e Cosmologia (Cosmo-Ufes) \& PPGFis, \\ Universidade Federal do Espírito Santo, Vitória, ES,  29075-910, Brazil}}
\newcommand{\INFNROMA}{\affiliation{INFN, Sezione di Roma, Piazzale Aldo Moro 2, 00185, Roma, Italy}}

\author{Mariano~Cadoni}
\email{mariano.cadoni@ca.infn.it}
\UniCa\INFNCa

\author{Leonardo~de Lima}
\email{leonardo.lima.88@edu.ufes.br}
\CosmoUfes

\author{Mirko~Pitzalis}
\email{mirko.pitzalis@ca.infn.it}
\UniCa\INFNCa

\author{Davi~C.~Rodrigues}
\email{davi.rodrigues@ufes.br}
\CosmoUfes

\author{Andrea~P.~Sanna}
\email{asanna@roma1.infn.it}
\INFNROMA

\title{
Cosmologically Coupled Black Holes with Regular Horizons
}
\begin{abstract}
    We present the most general and exact solution of Einstein's gravity sourced by an anisotropic fluid describing the cosmological embedding (CE) of a static and spherically-symmetric object, including black holes (BHs) or exotic compact objects, without radial energy influx and in an arbitrary Friedmann-Lema\^itre-Robertson-Walker (FLRW) cosmology. This is done fully considering backreaction of the local geometry on the cosmological dynamics. Our solution is free of curvature singularities at the would-be BH event horizon, thus solving a main issue of the CE of BH solutions proposed so far. As a byproduct,  we derive a new CE of the Schwarzschild BH - distinct from McVitties's original proposal - that is regular everywhere except  at the central singularity.
\end{abstract}

\maketitle

\section{Introduction }

It is an intriguing possibility that black holes (BHs) and other compact astrophysical objects may couple to the large-scale Friedmann-Lema\^itre-Robertson-Walker (FLRW) cosmological dynamics~\cite{McVittie:1933zz,Einstein:1945id,Einstein:1946zz,Croker:2024jfg,Faraoni:2023hin,Calza:2024qxn,Calza:2024xdh,Calza:2024fzo,dicke1964evolution,Vaidya:1968zza,DEath:1975jps,
Gautreau:1984pny,Cooperstock:1998ny,Nayak:2000mr,Baker:2000yh,Bolen:2000dz,Dominguez:2001it,Ellis:2001cq,Gao:2004cr,Sheehan:2004wa,
Nesseris:2004uj,Sultana:2005tp,Li:2006zh,Adkins:2006kw,McClure:2006kg,Sereno:2007tt,Faraoni:2007es,Balaguera-Antolinez:2007csw,Mashhoon:2007qm,Carrera:2008pi, Gao:2011tq,Faraoni:2014nba,Kopeikin:2014qna,Faraoni:2015saa,Mello:2016irl,Faraoni:2018xwo,Guariento:2019ock,Spengler:2021vxy,
Agatsuma:2022ewd,Croker:2019mup,Croker:2020,Croker:2020plg,Farrah:2023opk,Wang:2023aqe,Gaur:2023hmk,Parnovsky:2023wkc,Avelino:2023rac,Dahal:2023hzo,Gao:2023keg,Cadoni:2023lum,Cadoni:2023lqe,Cadoni:2024jxy,Cadoni:2024rri,BenAchour:2025vur,Asmanoglu:2025agc}. This idea, originally introduced by McVittie~\cite{McVittie:1933zz}, has recently regained attention within a genuinely dynamical setting. The cosmological embedding (CE) of BHs is usually constructed starting from a static, spherically-symmetric model and employing a set of physically motivated assumptions. It has been proposed that the mass $M$ of these cosmologically coupled (CC) astrophysical objects could vary with the redshift~\cite{Croker:2019mup,Cadoni:2023lum,Cadoni:2023lqe}, independently of accretion processes. This opened a new window to test fundamental properties of BHs and their mimickers, as well as the evolution of their mass spectrum at different redshifts~\cite{Cadoni:2023lum}.

General arguments based on causality and an effective fluid description suggest a power-law scaling of the mass with the scale factor $a$, $M \sim a^k$ with $-3 \le k \le 3$~\cite{Croker:2019mup}. These theoretical predictions have been tested against observations using both electromagnetic and gravitational-wave channels~\cite{Farrah:2023opk,Cadoni:2023lum,Rodriguez:2023gaa,Andrae:2023wge,Lei:2023mke,Amendola:2023ays,Lacy:2023kbb,Calza:2024qxn}, yielding, however, no conclusive evidence yet. The main issue at the observational level is the intrinsic difficulty in comparing homogeneous samples of BH mass spectra at different redshifts.

Setting aside observational challenges, the very existence of CC BH solutions raises significant theoretical issues. All CC BH models constructed so far feature a curvature singularity at the would-be event horizon of the corresponding static solution.

In the present work, we show that the theoretical difficulty mentioned above can be overcome. Inspired by the approach of~\cite{deLimaRodrigues2025,Cadoni:2023lum}, we construct solutions describing spherical objects embedded in a cosmological background with no radial matter flow (i.e., no accretion). We derive a general parametrization of the field equations in terms of a local mass function, together with the density and pressure components of the cosmological fluid. This parametrization is particularly useful being directly tied to physical intuition, as it employs the standard cosmological time coordinate, the comoving radius and a natural definition of the mass of the CE object.

Our construction is based on a given regular Misner-Sharp mass~\cite{Misner:1964je}, and is shown here to generate solutions that are regular everywhere, apart from the possible central singularity. As a consequence, when applied to BHs, their event horizons remain regular despite the embedding, thus solving the previously mentioned issue of CC BH models. We discuss both BHs with a central singularity and nonsingular BHs. Finally, we  derive a new CE of the Schwarzschild BH, distinct from the well-known McVittie solution~\cite{McVittie:1933zz}. \\

The structure of the paper is as follows. In \cref{sect:II}, we derive the general solution for the CE of local and spherically-symmetric objects and describe their physical properties.
In \cref{sect:III}, we specialize our discussion to BHs. Energy conditions are  analyzed in  \cref{sect:IV} and our conclusions are presented in \cref{sect:V} .\\

Throughout this work, we use units such that $c = G = 1$.

\section{Spacetime and the energy momentum tensor} 
\lb{sect:II}

The CE of a compact, spherically-symmetric and static objects in General Relativity can be described by the metric

\begin{equation}
\label{eq:LineElementMain}
    \dd s^2 = a^2(\eta) 
    \left[
        -e^{\alpha(\eta, r)} \, \dd\eta^2 
        + e^{\beta(\eta, r)} \, \dd r^2 
        + r^2 \, \dd\Omega^2
    \right] .
\end{equation}
Here, \( \alpha(\eta, r) \) and \( \beta(\eta, r) \) are metric functions depending on the conformal time \( \eta \) and \( r \), while \( a(\eta) \) is the scale factor.
The source of the gravitational field is an anisotropic fluid, which is described by the energy-momentum tensor (e.g., \cite{Herrera:1997plx, Ellis:1998ct, Horvat:2010xf} and references therein)
\begin{equation}
\label{eq:MomentTensorMain}
    T_{\mu \nu} 
    = (\rho + p_{\perp}) \, u_{\mu} u_{\nu} 
    + p_{\perp} \, g_{\mu \nu} 
    - (p_{\perp} - p_{\parallel}) \, w_{\mu} w_{\nu} \, ,
\end{equation}
where \( u_{\mu} \) and \( w_{\mu} \) satisfy the normalization conditions \( g^{\mu \nu} u_{\mu} u_{\nu} = -1 \), \( g^{\mu \nu} w_{\mu} w_{\nu} = 1 \), and \( u^{\mu} w_{\mu} = 0 \). Here, \( \rho \), \( p_{\parallel} \), and \( p_{\perp} \) denote the energy density and the radial and transverse pressures of the fluid, respectively. {All the three scalar quantities $\rho$, $p_\parallel$, and $p_\perp$, are at this point independent. The above $T_{\mu \nu}$ is the most general spherically-symmetric energy-momentum tensor without energy flux. As we will see below, $ \rho, \,  p_{\parallel}, \,  p_{\perp}$ are not dynamically independent, but they are constrained  by covariant conservation of the stress-energy tensor (see \cref{eq:rhoConservMain,eq:p} below). Notice also that anisotropy of the source in Einstein's  equation is achieved by taking  $p_{\perp}\neq p_{\parallel}$, which  generates an anisotropic stress, while the spacetime metric~\eqref{eq:LineElementMain} remains isotropic.}

The Einstein field equations following from the metric ansatz~\eqref{eq:LineElementMain} sourced by the energy–momentum tensor~\eqref{eq:MomentTensorMain} can be written as~\cite{Cadoni:2020jxe} {(see also Appendix \ref{app:field_eqs})}
\begin{subequations} \label{eq:FirstFieldEq}
\begin{align} 
& \frac{\dot{a}}{a} \alpha' + \frac{\dot{\beta}}{r} = 0\, ;\label{eq:FirstFieldEqa}\\
& 3 \frac{\dot{a}^2}{a^2} + \frac{e^{\alpha - \beta}}{r^2} \left(-1 + e^{\beta} + r\beta' \right) + \frac{\dot{a}}{a} \dot{\beta} = 8 \pi  e^{\alpha} a^2 \rho\, ;\label{eq:rhoMain} \\
& \frac{\dot{a}^2}{a^2} e^{\beta - \alpha} + \frac{1 - e^{\beta} + r\alpha'}{r^2} \nonumber \\
&+ e^{\beta - \alpha} \left(-2 \frac{\ddot{a}}{a} + \frac{\dot{a}}{a} \dot{\alpha} \right) = 8 \pi  e^{\beta} a^2 p_{\parallel}\, ; \label{eq:FirstFieldEqc}\\
\label{eq:rhoConservMain}
& \dot{\rho} + \frac{\dot{a}}{a} \left(3 \rho + 3 p_{\parallel} + r p'_{\parallel} \right) = 0\, ; \\
\label{eq:p}
& p'_{\parallel} + \frac{\alpha'}{2} \left(\rho + p_{\parallel} \right) + \frac{2}{r} \left(p_{\parallel} - p_{\perp} \right) = 0\, .
\end{align}
\end{subequations}
In the above, dot and prime refer to derivation with respect to $\eta$ and $r$, respectively.
The above equations are valid for any value of the spatial curvature parameter $\K$ (i.e., $\K=0$ or $\K=\pm 1$). For $\K\not=0$, $\rho$, $p_\parallel$ and $p_\perp$, depend on a geometric-contribution term that behaves like $\K a^{-2}$.

\subsection{Parametrization of the field equations in terms of local mass functions}

To solve the field equations~\eqref{eq:FirstFieldEq}, as a first step we decompose the energy density $\rho$ and the radial pressure $p_{\parallel}$ by isolating a purely homogeneous FLRW cosmological contribution (indicated by the subscript c) from the spherically symmetric local contribution (indicated by the subscript $l$)
\begin{equation}
\label{eq:ass}
\begin{aligned}
  \rho(r, \eta) &= \rho_{\text{c}}(\eta)+\rho_l(r, \eta), \\[4pt]
  p_{\parallel}(r, \eta) &= p_{\text{c}}(\eta) + p_{l \parallel}(r, \eta),
\end{aligned}
\end{equation}
where $\rho_{\text{c}}(\eta)$ and $p_{\text{c}}(\eta)$ satisfy the Friedmann cosmological equations

\begin{equation}
\label{eq:5}
\begin{aligned}
     &\frac{3 \dot{a}^2}{a^2} 
     = 8 \pi  a^2 \rho_{\text{c}}\, ; \\[6pt]
    &-2 \frac{\ddot{a}}{a} + \frac{\dot{a}^2}{a^2} 
     = 8 \pi  a^2 p_{\text{c}}\, ; \\[6pt]
   & \dot{\rho}_{\text{c}} + 3 \frac{\dot{a}}{a}
   \left( \rho_{\text{c}} +  p_{\text{c}} \right) = 0 \, .
\end{aligned}
\end{equation}
{In the above equations, the additional contribution from a possibly non-zero spatial curvature ($\K = \pm 1$) is included in $\rho_c$ and $p_c$.}

After some manipulation (explicit details can be found in \cref{Appendix: MS_parametrization}), the combination of Einstein's equation and covariant conservation of energy-momentum tensor leads to
\begin{subequations}
\label{eq:IntegrteFields0}
\begin{align}
&e^{-\beta} = 
1 - \frac{2\mms^{l}(r,\eta)}{ra}
+ \frac{8\pi}{3} a^2 r^2  \rho_{\text{c}}\left(e^{-\alpha} - 1\right)\,; \\
%
&\frac{\dot{a}}{a} \alpha' + \frac{\dot{\beta}}{r} = 0\, ;  \\
\label{IntegrteFields.cMain}
&\rho = 
\rho_{\text{c}} + \frac{1}{4\pi a^3 r^2} 
\partial_r \!\mms^{l}\, ; \\
\label{IntegrteFields.dMain}
&p_{\parallel} = 
p_{\text{c}} - \frac{1}{4\pi \dot{a} a^2 r^3} 
\partial_\eta \mms^{l}\, ; \\
%
&p_{\perp} = 
p_{\parallel} 
+ \frac{r}{2} \partial_r p_{\parallel} 
+ \frac{1}{4} r \alpha' \left(\rho + p_{\parallel}\right)\, .
\end{align}
\end{subequations}
The system of equations admits a uniquely-defined solution once the free function $\mms^{l}$ is specified, together with suitable boundary conditions. In particular, in order to retrieve {spatially-flat ($\K=0$)} FLRW cosmology at large radial distances, we must necessarily consider models satisfying $\alpha, \, \beta \to 0$ as $r\to \infty$. {For FLRW cosmologies with non-vanishing spatial curvature ($\K=\pm 1$), one imposes that $\alpha \to 0$ and $e^{\beta} \to 1/(1 - \K r^2)$ for sufficiently large $r$. 
In such limit, $\frac{2\mms^l}{ra}$ must reduce to $\K r^2$. From Eq.~\eqref{eq:IntegrteFields0}, one sees that $\rho \to \rho_c+\frac{3\K}{8\pi a^2}$, $p_\parallel \to p_c-\frac{\K}{8\pi a^2}$ and $p_\perp \to p_c-\frac{\K}{8\pi a^2}$ as long as $\alpha \to 0$ and
$e^{\beta} \to (1-\K r^2)^{-1}$. In the following sections, we shall restrict our analysis to the case $\K=0$, as this is the most relevant case for the late universe~\cite{Planck:2018vyg}.}

Regarding $\mms^{l}$, it physically represents the mass of the local object (MLO)
\begin{equation}\label{eq:ml}
    \mms^{l}(r,\eta)
    = \mms(r,\eta) - \frac{4\pi}{3}\,\ell^3 \rho_{\text{c}} \,,
\end{equation} 
which is defined in terms of the usual Misner-Sharp (MS) mass~\cite{Misner:1964je}
\begin{equation}
\label{eq:Ms_total}
    \mms(r,\eta)
    = \frac{\ell}{2}\left(1 - \nabla_a \ell \, \nabla^a \ell\right),
\end{equation}
where $\ell = a r$ is the areal radius. Therefore, $\mms^{l}$ is the MS mass of the compact object with the purely cosmological contribution subtracted. 
For compact objects, the $\rho_{\text{c}}$-dependent cosmological contribution can be neglected, giving $\mms^{l}\sim \mms$. Moreover, on short, non-cosmological, timescales, $\mms^{l}$ loses any dependence from $\eta$ and the mass is entirely determined  by the radial profiles  $\alpha(r)$ and $\beta(r)$ through \cref{eq:ml,eq:Ms_total}. 

\subsection{Exact  solutions}
\lb{sect:es}

Parametrizing the field equations in terms of the MLO is physically appealing and simplifies the form of the field equations. In general, however, once the functional form of the MLO is fixed, the full integration of the equations still requires solving nonlinear partial differential equations. Furthermore, the regularity of the MLO alone does not  guarantee the regularity of the corresponding metric solution.
A more convenient approach consists in parametrizing the solution in terms of a new function $\mathcal{M}(r,a)$, related to the MLO by the following definition 
\begin{align}
   \mathcal{M}(r,a) \equiv  & \mms^{l} (r,a) \,e^{-\theta } - \frac{\left(e^{-\theta }-1\right)}{2}\, \ell
\nonumber \\
&+ \frac{4\pi}{3}\,\rho_{\text{c}}\left(e^{-\theta }-1\right)\ell^3, 
\label{eq:alpha_beta}
\end{align}
where we have defined 
\begin{equation}
    \theta \equiv -\frac{1}{2}(\alpha + \beta)\, .
\end{equation}
In the following, we will also make use of another combination of the metric functions,
\begin{equation}
    \gamma\equiv -\frac{1}{2}(\alpha-\beta)\, .
\end{equation}
Using Eq.~\eqref{IntegrteFields.cMain}, we obtain
\begin{subequations}
\begin{align}
e^{-\gamma} & = \frac{1}{2} \left(f + \sqrt{f^{2} + \frac{32\pi}{3}\ell^{2}\rho_{\text{c}}}\right)\, ; \\
f &\equiv 1 - \frac{2 \mathcal{M}(r,a)}{ra} - \frac{8\pi}{3}\ell^{2}\rho_{\text{c}}\, .
\end{align}
\end{subequations}
In terms of $\gamma$ and $\theta $, the remaining equation, Eq.~\eqref{eq:FirstFieldEqa}, becomes
\begin{equation}\label{eq:g equation}
    -\frac{\dot{a}}{a}\,\gamma' + \frac{\dot{\gamma}}{r}
    = \frac{\dot{a}}{a}\,\theta ' + \frac{\dot{\theta }}{r}\,.
\end{equation}
To solve it, we introduce the coordinate transformation $(r,\eta)\to(\ell,a)$:
\begin{equation}\label{eq: transform}
 \partial_r \to a \,\partial_\ell,
 \qquad
 \frac{1}{\dot a}\partial_\eta \to  \partial_a + \frac{\ell}{a} \partial_\ell ,
\end{equation}

which leads to the equation
\begin{equation}
    \ell\,\partial_\ell \theta  + \frac{a}{2}\,\partial_a \theta 
    = \frac{1}{2}\,a\,\partial_a \gamma\, .
\end{equation}
The associated characteristic equations are 
\begin{align}
\label{Solution_eq_characteristics_methods}
\frac{\dd\ell}{\dd \tau} =\ell, \quad
\frac{\dd a}{\dd \tau}=\frac{a}{2}, \quad \frac{\dd\theta }{\dd \tau } & =\frac{1}{2}\,a\,\partial_a\gamma\left[\ell(\tau), \, a(\tau)\right]\, ,
\end{align}
 where $\tau$ parametrizes the characteristic curves. 

The solution of the first two equations yields
\begin{equation}
\ell(\tau)=\ell_0\,e^{\tau},\qquad
a(\tau)=a_0\,e^{\tau/2},
\end{equation}
with $\ell_0$ and $a_0$ integration constants.

It follows that $\ell/a^2 = \ell_0/a_0^2 = \mu$, so that each characteristic satisfies $\ell = \mu a^2$. Integrating along a characteristic from $a=a_0$ to $a_1$, we obtain
\begin{equation}
\theta (\ell,a_1)-\theta (\ell_*,a_0)
    = \int_{a_0}^{a_1}\partial_{a}\gamma(\mu \tilde a^2,\, \tilde a)\,\dd \tilde a,
\end{equation}
where $\ell_* = \mu a_0^{2}$. Substituting $\mu = \ell/a_1^{2}$ and $\ell_* = \ell\,a_0^{2}/a_1^{2}$, 
we arrive at the general form
\begin{equation}
\label{eq:general_g}
\theta (\ell,a_1)
 = \theta \!\left(\ell\,\frac{a_0^{2}}{a_1^{2}},a_0\right)
 + \int_{a_0}^{a_1}
   \partial_{a}\gamma\!\left(\frac{\ell}{a_1^{2}}\,\tilde a^{2},\, \tilde a\right)\,\dd \tilde a.
\end{equation}
Regrouping the equations, the complete system reads
\begin{subequations}
\label{eq:gensol}
\begin{align}
e^{-\gamma} &=  \frac{1}{2} \left(f + \sqrt{f^{2} + \frac{32\pi}{3}\ell^{2}\rho_{\text{c}}}\right)\, ;\\
f & \equiv 1 - \frac{2\mathcal{M}(r,a)}{ra} - \frac{8\pi}{3}\ell^2 \rho_{\text{c}}\, ; \\
\theta (\ell,a) &= \theta_0\left(\frac{\ell a_0^2}{a^2}\right)
    + \int_{a_0}^{a}
    \left.
    \frac{\partial}{\partial x}
    \gamma\!\left(\ell\frac{\tilde a^2}{a^2},x\right)
    \right|_{x=\tilde a}
    \dd \tilde a\, ,
\end{align}
\end{subequations}
where we defined $\theta_0(\ell) \equiv \theta(\ell, \, a_0)$. 

Note that the only remaining free parameter is $\mathcal{M}(r,a)$. This freedom is inherited from the MLO through the definition \eqref{eq:alpha_beta}, as already seen in the previous section. 

For any smooth choice of $\rho_{\text{c}}(a)$ and $\mathcal{M}(r,a)$, both $\theta$ and $\gamma$ are $C^2$ functions for $\ell, a > 0$. Thus, the resulting spacetime parametrized by \cref{eq:gensol} is free of curvature singularities, except at $\ell = 0$, representing the usual Big Bang singularity ($a = 0$), the usual BH spacelike singularity ($r = 0$) or both,  depending on the functional form of $\mathcal{M}(r,a)$.

It is also worth mentioning that the general solution given by \cref{eq:gensol} encodes both local object properties and cosmological dynamics. The dependence on $\rho_{\text{c}}$ represents  a dynamical interplay between the  large-scale cosmological  dynamics and the local geometry. Crucially, it is the presence of $\rho_{\text{c}}$ in \cref{eq:gensol} that allows to avoid curvature singularities at the would-be event horizon, which would otherwise be present in the solutions for $f=0$.

Let us conclude the section with some comments on the physical properties of $\mathcal{M}(r,a)$. Although having been initially introduced merely as a tool to simplify the integration of the field equations, it acquires a clear physical meaning when evaluated at the apparent horizon. For spherically-symmetric metrics, the location of the latter is determined by the condition~\cite{Faraoni_2009, Faraoni2015-ls} 
\begin{align}
    \nabla_{\mu}\ell \,\nabla^{\mu}\ell
    &= 1 - \frac{2\mms(r_\text{h},a)}{\ell}\nonumber \\
    &= 1 - \frac{2\mms^{\,l}(r_\text{h},a)}{\ell}
      - \frac{8\pi}{3}\, r_\text{h}^2 a^2 \rho_{\text{c}}
    = 0 \, ,
    \label{eq:MSl1}
\end{align}
where $r_\text{h} = r_\text{h}(\eta)$ is the radius of the apparent horizon.

Using Eq.~\eqref{eq:alpha_beta}, we find 
\begin{equation}
    1 - \frac{2\mathcal{M}(r_\text{h},a)}{\ell}
      - \frac{8\pi}{3} r_\text{h}^2 a^2 \rho_{\text{c}}
    = 0 \, ,
    \label{eq:MSl2}
\end{equation}
which reveals that, at the apparent horizon, 
\begin{equation}
    \mms^{l}(r_\text{h}, a) = \mathcal{M}(r_\text{h}, a)\, .
    \label{eq:EquivalenceMasses}
\end{equation}
Finally, $\mathcal{M}(r,a)$ is related to the MS mass at the apparent horizon by
\begin{equation}\label{eq:M}
    \mms(r_\text{h},a)
    = \mathcal{M}(r_\text{h}, a)
      + \frac{4\pi}{3} r_\text{h}^{3} a^{3} \rho_{\text{c}}\, e^{\theta }\, .
\end{equation}
When the cosmological contributions can be neglected, this equation automatically reduces to \cref{eq:EquivalenceMasses}.

\section{Cosmologically embedded black holes}
\lb{sect:III}

Let us now focus on the subclass of solutions describing the CE of BHs, including both the singular Schwarzschild and nonsingular BH models, like those analyzed in Ref.~\cite{Cadoni:2022chn}. The simplest way to contruct CE of BH solutions is to start from a mass profile $M(r)$ describing a static, spherically-symmetric BH and promote the radial coordinate $r\to ra=\ell$. This method has been firstly introduced by McVittie, working however in isotropic coordinates~\cite{McVittie:1933zz}. Notice that, since after the embedding, $\alpha$ and $\beta$ will be functions of $\ell$ only, the condition $\alpha=-\beta$, immediately follows from \cref{eq:FirstFieldEqa},  ensuring the absence of radial energy flux. In this way, we can construct two different classes of exact solutions. Despite the fact that the first one can be obtained as a particular case of \cref{eq:gensol}, it is easier to derive it directly from Einstein's field equations
\begin{subequations}
\label{eq:exactsol1}
\begin{align} 
        e^{\alpha} & =e^{-\beta}= 1 - \frac{2M(\ell)}{\ell}\, ;\\
        \rho &= e^{-\alpha}\rho_{\text{c}}\left[ 1-\frac{k(\ell)}{3}\right]+ \frac{ \partial_\ell M}{4\pi  \ell^2}\, ;\\
        p_{\parallel} &= e^{-\alpha} \left[ p_{\text{c}}+\frac{k(\ell)}{3}\rho_{\text{c}}\right]- \frac{ \partial_\ell M}{4\pi  \ell^2}\, ;\\
        p_{\perp} &= p_{\parallel} + \frac{\ell}{2} \partial_\ell p_{\parallel} + \frac{k(\ell)}{4} \left(\rho + p_{\parallel} \right)\, ,
\end{align}
\end{subequations}
where $k\equiv r \partial_r\alpha=\ell \partial_\ell\alpha$.
Differently from $\mms$, the mass function $M(\ell)$ is entirely determined by the local object, as it does not contain any $\rho_{\text{c}}$-dependent cosmological contribution.
$M(\ell)$ is related to the MS mass through
\begin{equation}
\mms = M(\ell)+ \frac{4\pi}{3} \rho_{\text{c}} \,\ell^3 \left[ 1 - \frac{2M(\ell)}{\ell}\right]^{-1}\, .
\end{equation}
Depending on the functional form of $M(\ell)$, our solution describes the CE of well-known spherically-symmetric BHs. 

 A first physically relevant case is obtained by choosing $M(\ell)=M_0= \text{const.}$, which corresponds to the CE of the Schwarzschild BH. However, the resulting spacetime differs from the McVittie solution~\cite{McVittie:1933zz}, as the two cannot be related by any coordinate transformation. This inequivalence arises because the McVittie spacetime is derived under the assumption of vanishing energy flux in the isotropic radial coordinate and is sourced by an isotropic fluid. In contrast, our solutions enforce the no-flux condition in the comoving areal coordinate and are sourced by an anisotropic fluid.
When expressed in terms of the proper length $\ell$, our solution reads
\begin{equation}
\begin{split}
\dd s^2 =&- \left[ 1 -\frac{2M_0}{\ell} - \left(1 -\frac{2M_0}{\ell}\right)^{-1} \frac {\ell^2}{a^2} H^2 \right] \dd t^2 \\
&+ \left(1 -\frac{2M_0}{\ell}\right)^{-1} \dd\ell^2 \\
&-2 \left(1 -\frac{2M_0}{\ell}\right)^{-1} H\ell \,\dd t \,\dd\ell +\ell^2 \dd\Omega^2\, ,
\end{split}
\end{equation}
with $t$ the cosmological time. Interestingly, it bears some similarities with the Sultana-Dyer solution~\cite{SultanaDyer2005,Faraoni_2009}, which, however, describes the cosmological embedding of a  BH with non-zero energy fluxes. 

As discussed in the previous section, the absence of curvature singularities in the CC solution is a direct consequence of backreaction effects, encoded in the $\rho_{\text{c}}$-dependent term appearing in the metric functions. For the class of solutions derived from \cref{eq:exactsol1}, however, this mechanism is absent, since the information associated with $\rho_{\text{c}}$ is not geometrically encoded. It instead resides entirely in the matter source. Consequently, we expect the spacetime to develop a curvature singularity at $\ell_\text{h}$, corresponding to the location of the static BH event horizon. This radius $\ell_\text{h}$ is determined by the condition $2M(\ell_\text{h})=  \ell_\text{h}$. This expectation can be explicitly confirmed by evaluating the curvature invariants of the spacetime described by~\eqref{eq:exactsol1}.

In view of this pathology, our aim is now to construct a  second class of solutions, which are free from this singularity at the horizon. We will show that  these regular solutions can be easily obtained using the exact solution discussed in \cref{sect:es}.

Given that we recover the static BH solution with a mass profile $M(r)$ in the limit $\rho_{\text{c}} \to 0$, it is sufficient to identify $\mathcal{M}(\ell)=M(\ell)$ in \cref{eq:gensol}. We thus obtain
\begin{subequations}
\lb{rh}
\begin{align}
e^{-\gamma} 
    & = \frac{1}{2}\left(\,f + \sqrt{f^2 + \frac{32\pi}{3}\ell^2\rho_{\text{c}}}\right)\, ; \\
 f &\equiv 1 - \frac{2M(\ell)}{\ell} 
      - \frac{8\pi}{3}\ell^2 \rho_{\text{c}}\, ; \\
\theta (\ell,a) & \label{eq:gensol_b}
    = \theta_0\!\left(\frac{\ell a_0^2}{a^2}\right)
    + \int_{a_0}^{a}
    \left.
    \frac{\partial}{\partial x}
    \gamma\!\left(\ell\frac{\tilde a^2}{a^2},x\right)
    \right|_{x=\tilde a}
    \dd \tilde a\, .
\end{align}
\end{subequations}
This solution is now free of curvature singularities, and outside the apparent horizon we have $\partial_a \gamma(\ell,a) \sim \partial_a \rho_{\text{c}} \approx 0$, so that the contribution coming from the integral in \cref{eq:gensol_b} becomes negligible. This is a quite important result and confirms the findings above: the horizon singularity - observed in all previously studied BH embeddings - is resolved once one considers solutions that consistently incorporate the dynamical interplay between   the local metric and the large-scale cosmological dynamics. Therefore, the curvature singularity at the horizon, featured by the McVittie solution (see, e.g., Ref.~\cite{Modesto:2025cre}), is here cured. 

We also notice that this second class of solution reduces to the former \eqref{eq:exactsol1} outside the apparent horizon. Indeed, by fixing $\theta_0(\ell)=0$, and working to first order in $\dot a/a$, the solution outside the apparent horizon is
\begin{align}
    \dd s^2 \approx \, & a^2(\eta) 
    \biggl[
        -\left(1-\frac{2M(\ell)}{\ell}\right)\! \dd\eta^2 \nonumber\\
        &+ \dd r^2 \left(1-\frac{2M(\ell)}{\ell}\right)^{-1}
        + r^2\, \dd\Omega^2
    \biggr],
\end{align}
which coincides with the metric components in Eq.~\eqref{eq:exactsol1}. However, close to the apparent horizon, the contribution coming from the integral in \cref{eq:gensol_b} becomes dominant and  grows as $a/a_{0}$ increases, leading to a genuinely different geometry. 

The results presented so far naturally extend to the CE of nonsingular BHs with de Sitter core (see, e.g., Ref.~\cite{Cadoni:2023lum}). They are characterized by a mass function $M(r)$ behaving as $r^3$ near $r=0$ and as $M_0$ as $r\to\infty$. Although a possible CE for such models has already been proposed in literature~\cite{Cadoni:2023lum,Cadoni:2023lqe,Cadoni:2024rri}, these embeddings still exhibit a curvature singularity at the horizon\footnote{This feature has not been previously emphasized, but it can be readily verified by computing the curvature invariants of the solutions presented in Ref.~\cite{Cadoni:2023lqe}.}. Remarkably, the embedding introduced in the present work removes this pathology and the horizon singularities appearing in the CE of otherwise regular BH models are eliminated. 

Finally, our results provide further insight into the theorem of Ref.~\cite{Faraoni:2024ghi}, which forbids the existence of regular static horizons in any CE of black hole solution. The present construction is fully consistent with this theorem, since the solutions obtained here do not admit static event horizons, but only apparent horizons. At the same time, our results explain the singular behaviour at the horizon of the solution \eqref{eq:exactsol1} as a direct consequence of neglecting the, $\rho_{\text{c}}$-dependent contribution.

\section{Energy conditions}
\lb{sect:IV}
When interpreted as effective models within General Relativity, one generally expects the fluid sourcing the CC BH solutions to be given by some kind of exotic matter. As a consequence, the standard energy conditions—such as the weak (WEC), null (NEC), and dominant (DEC) energy conditions—may be violated in certain  spacetime regions. In particular, as already observed for the McVittie’s solution, violations of the energy conditions can occur in the vicinity of the horizon. To investigate these conditions, it is physically reasonable to assume that the cosmological fluid characterized by $\rho_{\text{c}}$ and $p_{\text{c}}$ satisfies all of them, namely $\rho_{\text{c}} \ge 0$, $\rho_{\text{c}} + p_{\text{c}} \ge 0$, and $\rho_{\text{c}} \ge |p_{\text{c}}|$.

We first consider the CC solutions $\eqref{eq:gensol}$ and assume that terms proportional to derivatives of $\mms^{l}$ can be neglected. This assumption is justified, for instance, when $\mms^{l} = M_0 = \text{const.}$, and more generally holds for models in which the density profile decays sufficiently rapidly outside the BH. We get, {using the expressions  for  $\rho, \, p_{\parallel},\,p_{\perp}$  given by \eqref{eq:IntegrteFields0},  }
\begin{equation}
\begin{split}
    \rho &=\rho_{\text{c}}\,, \\
    \rho+p_{\parallel} & =\rho_{\text{c}}+p_{\text{c}}\,,\\  
    \rho+p_{\perp}&=\left(1+\frac{k}{4}\right)\left(\rho_{\text{c}}+p_{\text{c}}\right)\,,
\end{split}    
\end{equation}
with $k$ defined as in \cref{eq:exactsol1}. Since $k>0$, the WEC, and hence the NEC, are always satisfied outside the BH.  Conversely, considering an equation of state $p_{\text{c}}= \omega \rho_{\text{c}}$ for the cosmological background, we find that for $\omega \ge 0$ the DEC is always satisfied in the asymptotically-flat region 
(corresponding to $k=0$), but is typically violated in the near-horizon region. For $\omega<0$ the opposite behavior occurs.

Let us now turn to the class of solutions \eqref{eq:exactsol1}. In this case one finds {using the expressions  for  $\rho, \, p_{\parallel},\,p_{\perp}$  given by \cref{eq:exactsol1}}, $e^{\alpha}(p_{\parallel}+\rho)= p_{\text{c}}+\rho_{\text{c}}.$
The NEC is, therefore, always satisfied in the BH exterior. 
On the other hand, $\rho + p_\perp$ depends on $k$ and its $r$-derivative. In general, it will always exist a value  $\ell_0$, typically of the order of $\ell_\text{h}$ such that for $\ell\le \ell_0$, the inequality $\rho + p_\perp\ge 0$ is violated. Thus, we expect the NEC (hence the WEC) to be violated in the near-horizon region. A similar argument shows that this is also true for the DEC.

We conclude by noting that the inclusion of the backreaction term in \cref{eq:gensol} plays a crucial role. Beyond curing the curvature singularity at the horizon of the solutions \eqref{eq:exactsol1}, it also restores the validity of the WEC in the BH exterior.

\section{Conclusions}
\lb{sect:V} 

In this paper, we have derived the general solution describing the CE of local compact and spherically-symmetric objects, including BHs, within an extended framework in which the sources are modeled by anisotropic fluids. These solutions consistently incorporate large-scale cosmological dynamics directly into the spacetime geometry through a backreaction term, encoded in the explicit dependence of the metric functions on the cosmological energy density $\rho_{\text{c}}$. We have shown that the inclusion of this contribution removes the spacetime singularities that would otherwise appear at the would-be static event horizon, thereby resolving one of the main shortcomings of previously proposed CC solutions. Moreover, this generalized framework enables the construction of new classes of CC BHs. In particular, we have identified a novel cosmological embedding of the Schwarzschild black hole, distinct from the original solution proposed by McVittie.  

Our results also elucidate the physical content of the theorem proved in Ref.~\cite{Faraoni:2024ghi}, which forbids the existence of regular static horizons in CE solutions. On the one hand, our general CE solution fully complies with the theorem, as it admits an apparent, but not a static horizon. On the other hand, our analysis clarifies the fate of the event horizon of a local static BH under cosmological embedding. Prior to the embedding, the BH event horizon, defined by the condition $2M(r_\text{H})=r_\text{H}$, is a completely regular surface. If the CE is performed while neglecting backreaction, namely by employing \cref{eq:exactsol1}, the resulting spacetime develops a curvature singularity at the surface defined by $2M(\ell_\text{h}) = \ell_\text{h}$. This singular behavior is avoided once backreaction effects are properly included, that is, when the CE is constructed using \cref{rh}.

{This work is not the first to investigate cosmological coupling and use a mass-related function to parametrize possible GR solutions, as we did in Eqs.~(\ref{eq:IntegrteFields0}, \ref{eq:gensol}) and as previously considered in Ref.~\cite{Nandra:2011ug}. However, the analysis in Ref.~\cite{Nandra:2011ug} assumes isotropic pressure, which prevents the avoidance of singularities beyond the central one for $\K=0$~\cite{Nolan:1998xs, Nolan:1999wf}. In contrast, we have developed a new parametrization of GR solutions that relaxes the assumption of isotropic pressure. We have shown that this generalized framework can eliminate all singularities—apart from the central one—even in the case $\K=0$. Consequently, our results provide a partial extension of the $\K=0$ solutions of Ref.~\cite{Nandra:2011ug} to the case of an anisotropic fluid.} 

An important issue concerns the validity of the energy conditions for the solutions obtained in this framework. While the WEC is satisfied in the BH exterior, for a generic equation of state of the cosmological fluid, the DEC is violated in certain spacetime regions, typically near the horizon. Violations of the DEC are commonly associated with exotic matter and superluminal energy transport. Although such features are clearly undesirable, they do not appear to represent a fatal shortcoming of the present model. In our framework, the anisotropic fluid sourcing the CE solution should be interpreted as an effective, coarse-grained description of inhomogeneities embedded within an otherwise homogeneous and isotropic universe. As a result, a violation of the DEC at the level of the effective fluid does not necessarily imply a corresponding violation at the fundamental, microscopic level. An interesting open question concerns whether, in the case $\omega\ge 0$, violations of the DEC occur inside or outside the apparent horizon. Since the latter is always located outside the BH event horizon, it is conceivable that DEC violations may arise outside the event horizon while remaining confined within the apparent one.

\acknowledgements
LL thanks \textit{Fundação de Amparo à Pesquisa e Inovação do Espírito Santo} (FAPES, Brazil) for support.
DCR thanks CNPq (Brazil) and FAPES (Brazil) for partial support. APS is supported by the MUR FIS2 Advanced Grant ET-NOW (CUP:~B53C25001080001) and by the INFN TEONGRAV initiative. He also gratefully acknowledges support from a research grant funded under the INFN–ASPAL agreement as part of the Einstein Telescope training program.

\appendix

\section{{Detailed derivation of the field equations}}\label{app:field_eqs}

The field equations \eqref{eq:FirstFieldEq} were previously derived using similar notation and assumptions  in Refs.~\cite{Cadoni:2020jxe, Cadoni:2023lum, Cadoni:2024jxy}. For completeness and clarity, we present the complete derivation here. A \textsc{Mathematica} notebook is also provided \footnote{\href{github.com/davi-rodrigues/ccbh-rh}{https://github.com/davi-rodrigues/ccbh-rh}.} which allows to reproduce the following results. For brevity, we use the index notation $(t, r, \theta, \phi) \leftrightarrow (0, 1, 2, 3)$.

The non-vanishing Christoffel symbols read
\begin{subequations} 
\begin{align}
\Gamma^0_{00} &= \frac{\dot a}{a} + \frac12 \dot\alpha ,
\qquad \Gamma^0_{01} = \frac12 \alpha' ,
\\[0.5em]
\Gamma^0_{11} &= \frac{e^{-\alpha+\beta}}{2a}\left(2\dot a + a\dot\beta\right) ,
\\[0.5em]
\Gamma^0_{22} &= \frac{e^{-\alpha} r^2 \dot a}{a} ,
\qquad
\Gamma^0_{33} = \frac{e^{-\alpha} r^2 \sin^2\theta\, \dot a}{a} ,
\\[0.5em]
\Gamma^1_{01} & = \frac{\dot a}{a} + \frac12 \dot\beta ,
\qquad
\Gamma^2_{02}  = \Gamma^3_{03} = \frac{\dot a}{a} ,
\\[0.5em]
\Gamma^2_{12} &= \Gamma^3_{13} = \frac{1}{r} ,
\qquad \Gamma^3_{23} = \cot\theta ,
\\[1em]
\Gamma^1_{00} &= \frac12 e^{\alpha-\beta}\alpha' ,
\qquad \Gamma^1_{11} = \frac12 \beta' ,
\\[0.5em]
\Gamma^1_{22} &= - e^{-\beta} r ,
\qquad
\Gamma^1_{33} = - e^{-\beta} r \sin^2\theta ,
\\[0.5em]
\Gamma^2_{33} &= -\cos\theta \sin\theta .
\end{align}
\end{subequations} 

The Einstein tensor components are
\begin{subequations} 
\begin{align}
G^{0}_{0} &= -\frac{e^{-\alpha}}{a^{2}}
\Bigg[
\frac{3\dot a^{2}}{a^{2}}
+ \frac{e^{\alpha-\beta}}{r^{2}}\Big(-1+e^{\beta}+r\beta'\Big)
+ \frac{\dot a}{a}\dot\beta
\Bigg], \\[0.5 cm]
G^{0}_{1} &= -\frac{e^{-\alpha}}{a^{3}r}
\Big(r\dot a\,\alpha' + a\dot\beta\Big), \\[0.5 cm]
G^{1}_{1} &= \frac{1}{a^{4}}
\Bigg[
e^{-\alpha}\dot a^{2}
+ \frac{a^{2}e^{-\beta}}{r^{2}}\Big(1-e^{\beta}+r\alpha'\Big)
+ \nonumber \\ 
& +  a e^{-\alpha}\Big(-2\ddot a+\dot a\,\dot\alpha\Big)
\Bigg], \\[0.5 cm]
G^{2}_{2} &= G^{3}_{3} = \frac{e^{-\alpha-\beta}}{4a^{4}r}
\Bigg\{ 4e^{\beta}r\dot a^{2} + 4ae^{\beta}r\dot a\Big(\dot\alpha-\dot\beta\Big)
\nonumber\\
& + a\Bigg[-8e^{\beta}r\ddot a
+ a\Big[
e^{\alpha}r \alpha'^{2}
-2e^{\alpha}\beta'
- e^{\alpha}\alpha'(-2+r\beta')
\nonumber\\
&+ 2e^{\alpha}r\alpha''
+ e^{\beta}r\left(\dot\alpha\dot\beta-\dot\beta^{2}-2\ddot\beta\right)
\Big]\Bigg]\Bigg\}.
\end{align}
\end{subequations} 

The energy-momentum tensor is defined in Eq.~\eqref{eq:MomentTensorMain}. Using {an appropriate comoving reference frame in which $u^\mu=(u^0,0,0,0)$ and $w^\mu=(0,w^1,0,0)$,} the energy-momentum tensor components are
\begin{equation}
    T_{\mu}^\nu= \mbox{diag} \begin{pmatrix}
    -\rho,  & p_\parallel, & p_\perp, & p_\perp
    \end{pmatrix}\, .
\end{equation}
Using the Einstein field equations and from $T_0^1 = 0$ (absence of fluxes), one finds Eq.~\eqref{eq:FirstFieldEqa}. Equations \eqref{eq:rhoMain} and \eqref{eq:FirstFieldEqc} are directly derived from $G_0^0 = - 8 \pi \rho$ and $G_1^1 =  8 \pi p_\parallel$. To find the remaining equations, it is useful to consider the energy-momentum tensor conservation, which leads to the two independent equations
\begin{subequations}
\begin{align}
    \nabla^\mu T_{\mu 0} & = -\dot \rho - \frac{\dot a}{a} \left( 3 \rho + p_\parallel + 2 p_\perp \right)  - \frac{\dot \beta}{2} (\rho + p_\parallel)  \, , \label{eq:DivT0}\\[0.5cm]
    \nabla^\mu T_{\mu 1} & =  p_\parallel' + 2 \frac{p_\parallel - p_\perp}{r} + \frac{\alpha'}{2}(\rho + p_\parallel)  \, .  \label{eq:DivT1}
\end{align}
\end{subequations}

Equation \eqref{eq:p} comes straight from Eq.~\eqref{eq:DivT1}. At last, to find Eq.~\eqref{eq:rhoConservMain}, one uses Eqs.~\eqref{eq:FirstFieldEqa} and \eqref{eq:DivT1} to replace $\dot \beta$ from Eq.~\eqref{eq:DivT0}.

\section{The Misner-Sharp mass parametrization}\label{Appendix: MS_parametrization}

Expressing \( p_{\perp} \) in terms of \( \rho \), \( \alpha \) and \( p_{\parallel} \) by using \cref{eq:p}, the system of the field equations \eqref{eq:FirstFieldEq} can be recast as:
\begin{subequations}\label{eq:FieldEquations}
\begin{align}
\label{eq:FieldEquations.a}
& \frac{\dot{a}}{a} \alpha' + \frac{\dot{\beta}}{r} = 0\, ; \\
\label{eq:FieldEquations.b}
& 3 \frac{\dot{a}^2}{a^2} + \frac{e^{\alpha - \beta}}{r^2} \left(-1 + e^{\beta} + r\beta' \right) + \frac{\dot{a}}{a} \dot{\beta} 
= 8 \pi  e^{\alpha} a^2 \rho\, ; \\
\label{eq:FieldEquations.c}
& \frac{\dot{a}^2}{a^2} e^{\beta - \alpha} + \frac{1 - e^{\beta} + r\alpha'}{r^2} 
\nonumber \\
&+ e^{\beta - \alpha} \left(-2 \frac{\ddot{a}}{a} + \frac{\dot{a}}{a} \dot{\alpha} \right) = 0\, ;\\
\label{eq:FieldEquations.d}
& \dot{\rho} + \frac{\dot{a}}{a} \left(3 \rho + 3 p_{\parallel} + r p'_{\parallel} \right) = 0\, .
\end{align}
\end{subequations}
Inserting Eq.~\eqref{eq:ass} into the system \eqref{eq:FieldEquations}, we get
\begin{subequations}
\begin{align}
\label{eq:8.a}
& \frac{\dot{a}}{a} \alpha' + \frac{\dot{\beta}}{r} = 0\, ; \\
\label{eq:8.b}
& \rho_l = \frac{1}{8\pi a^2 r^2} \partial_r\!\left[r\left(1 - e^{-\beta}\right)\right]
\nonumber \\
&- \frac{\rho_{\text{c}}}{3r^2} \partial_r\!\left[r^3\left(1 - e^{-\alpha}\right)\right]\, ; \\
\label{eq:8.c}
& p_{l \parallel } = \frac{1}{8\pi a^2 r^2}\!\left[e^{-\beta}(1+k) - 1\right]
+ \frac{\dot{a}}{8\pi a^3} e^{-\alpha}\dot{\alpha}\nonumber \\
&- p_{\text{c}}\left(1 - e^{-\alpha}\right)\, ; \\
\label{eq:8.d}
& \dot{\rho}_l + \frac{\dot{a}}{a}\!\left(3\rho_l + 3p_{l\parallel } + r p'_{l \parallel}\right)=0\, ,
\end{align}
\end{subequations}
where \( k(r,\eta) \equiv r\,\partial_r \alpha \). \Cref{eq:8.b} can be integrated with respect to \( r \), leading to
\begin{align}
\label{eq:9}
\frac{I(r,\eta)}{a^3} = &\frac{r(1 - e^{-\beta})}{2a^2}
- \frac{4\pi\rho_{\text{c}}}{3} r^3(1 - e^{-\alpha})- F(\eta)\, ,
\end{align}
where $I(r,\eta) \equiv 4\pi a^3 \int_0^r \rho_l(u,\eta)\, u^2\, \dd u$ and \( F(\eta) \) is an integration function.

From \cref{eq:8.a}, we obtain
\begin{equation}
    e^{-\beta}(1+k) = \frac{1}{\dot{a}} \partial_\eta\left(e^{-\beta} a\right).
\end{equation}

Thus, \cref{eq:8.c} becomes
\begin{align}
p_{l \parallel} = &\frac{1}{8\pi a^2 r^2}\!\left[\frac{1}{\dot{a}}\partial_\eta\left(e^{-\beta} a\right) - 1\right]
- \frac{\dot{a}}{8\pi a^3}\partial_\eta e^{-\alpha}\nonumber \\ 
&- p_{\text{c}}\left(1-e^{-\alpha}\right).
\end{align}

From this we obtain
\begin{equation}
\begin{split}
\partial_\eta\!\left(e^{-\beta} a\right) =\;& 
\dot{a} 
+ r^2 \frac{\dot{a}^2}{a} \partial_\eta e^{-\alpha}
+ 8\pi r^2 \dot{a} a^2 p_{l \parallel} \\
& + 8\pi r^2 \dot{a} a^2 p_{\text{c}}(1 - e^{-\alpha})\, .
\end{split}
\end{equation}

Integrating the previous equation we get
\begin{equation}\label{eq:13}
\begin{split}
e^{-\beta} =\;& 1 
+ \frac{r^2}{a} \int \frac{\dot{a}^2}{a} \partial_\eta e^{-\alpha}\, \dd\eta
+ \frac{8 \pi r^2}{a} \int \dot{a} a^2 p_{l \parallel}\, \dd\eta \\
& + \frac{8 \pi r^2}{a} \int \dot{a} a^2 p_{\text{c}}(1-e^{-\alpha})\, \dd\eta
+ \frac{h(r)}{a},
\end{split}
\end{equation}
with $h(r)$ an integration function.

Using \cref{eq:FieldEquations.d}, we can write
\begin{equation}
    \frac{1}{a^2}\partial_\eta(a^3\rho_l) 
= -\frac{\dot{a}}{r^2} \partial_r(r^3 p_{l\parallel }).
\end{equation}

Integrating  the previous equation in \( r \) we get 
\begin{equation}\label{eq:m_p}
\partial_\eta I (r,\eta) = -4 \pi \dot{a} a^2 r^3 p_{l \parallel} - \partial_\eta \tilde F (\eta),
\end{equation}
with $\tilde F$ an integration function.

We then obtain from \cref{eq:13}
\begin{equation}\label{eq:17}
\begin{split}
e^{-\beta} =\;& 1 - \frac{2I(r,\eta)}{ra} - \frac{2\tilde F(\eta)}{ra}
+ \frac{r^2}{a}\!\int \frac{\dot{a}^2}{a}\partial_\eta e^{-\alpha}\, \dd\eta \\
& + \frac{8\pi r^2}{a} \int \dot{a} a^2 p_{\text{c}}(1-e^{-\alpha})\, \dd\eta
+ \frac{h(r)}{a}.
\end{split}
\end{equation}

Observe that
\begin{equation}
\int \frac{\dot{a}^2}{a} \, \partial_\eta e^{-\alpha} \, \dd\eta
= \frac{\dot{a}^2}{a} e^{-\alpha}
- \int \partial_\eta \!\left(\frac{\dot{a}^2}{a} \right) e^{-\alpha} \, \dd\eta.
\end{equation}

Comparing with \eqref{eq:9}, we obtain the consistency conditions
\begin{equation}
    h(r)=0, \qquad \tilde F(\eta)= F(\eta),
\end{equation}
and thus
\begin{align}
\label{eq:22_final}
e^{-\beta} =& 1 - \frac{\left[2I(r,\eta)+F(\eta)\right]}{ra}  \nonumber\\
&+\frac{8\pi}{3} r^2  a^2 \rho_{\text{c}} \left(e^{-\alpha}-1\right).
\end{align}

From the definition \eqref{eq:ml}, $I + \frac{1}{2}F(\eta)$ corresponds to the MLO, $\mms^{l}$. Therefore, the final system can be written as
\begin{subequations}
\label{eq:IntegrteFields}
\begin{align}
\label{IntegrteFields.a}
&e^{-\beta} = 
1 - \frac{2\mms^{l}(r,\eta)}{ra} 
+ \frac{8\pi}{3} r^2 a^2 \rho_{\text{c}} \left(e^{-\alpha} - 1\right)\, ; \\
\label{IntegrteFields.b}
&\frac{\dot{a}}{a} \alpha' + \frac{\dot{\beta}}{r} = 0\, ; \\
\label{IntegrteFields.c}
&\rho = 
\rho_{\text{c}} + \frac{1}{4\pi a^3 r^2} 
\partial_r \!\mms^{l}\, ; \\
\label{IntegrteFields.d}
&p_{\parallel} = 
p_{\text{c}} - \frac{1}{4\pi \dot{a} a^2 r^3} 
\partial_\eta \mms^{l}\, ; \\
\label{IntegrteFields.e}
&p_{\perp} = 
p_{\parallel} 
+ \frac{r}{2} \partial_r p_{\parallel} 
+ \frac{1}{4} r \alpha' \left(\rho + p_{\parallel}\right) \, .
\end{align}
\end{subequations}
Once the cosmological background is fixed, the only free function in the system is the MLO mass.

\end{document}